\documentclass{aa}  
\usepackage{graphicx}
\usepackage{txfonts}
\begin{document}
\def\kms{km s$^{-1}$}
\def\etal{et al.}
\def\hi{H\,{\sc i}}
\def\hii{H\,{\sc ii}}
\def\deg{$^\circ$}
\def\msun{M$_\odot$}
\def\mjyb{mJy beam$^{-1}$}
\def\jyb{Jy beam$^{-1}$}
\def\msunyr{M$_\odot$ yr$^{-1}$}
\def\cmtres{cm$^{-3}$}
\def\cmdos{cm$^{-2}$}
\def\ojo{\fbox{\bf !`$\odot$j$\odot$!}}
\def\por{$\times$}
\def\radec{$R.A.,Decl.(J2000)$}

\title{Unveiling the molecular environment of the ring nebula RCW\,78}

\author{C.E. Cappa\inst{1,2}\thanks{Member of 
Carrera del Investigador, CONICET, Argentina},
M. Rubio\inst{3},
M.C. Mart\'{\i}n\inst{2*}
\and
G.A. Romero\inst{1,2,4}
}

\offprints{C.E. Cappa}

\institute{Facultad de Ciencias Astron\'omicas y Geof\'{\i}sicas,  Universidad
Nacional de La Plata, Paseo del Bosque s/n, 1900 La Plata, Argentina\\
\email{ccappa@fcaglp.fcaglp.unlp.edu.ar}
\and
Instituto Argentino de Radioastronom\'{\i}a, C.C. 5, 1894 Villa Elisa, 
Argentina
\and
Departamento de Astronom\'{\i}a, Universidad de Chile, Casilla 36-D, 
Santiago, Chile 
\and
Departamento de F\'{\i}sica y Astronom\'{\i}a, Facultad de Ciencias,
Universidad de Valpara\'{\i}so, Chile  \\
}

\date{Received September 15, 1996; accepted March 16, 1997}

 
 \abstract
 {}
 {We present a study of the ionized, neutral atomic, and molecular  gas 
associated with the ring nebula RCW\,78 around the WR star 
HD\,117688~(= WR\,55) with the aim of analyzing the distribution of the 
associated gas  and  investigating its energetics.
}
   {We based our study on $^{12}$CO(1-0) and $^{12}$CO(2-1) observations of 
the brightest section of the nebula carried out with the SEST telescope with 
angular resolutions of 45\arcsec\ and 22\arcsec, respectively; and on  
complementary $^{12}$CO(1-0) data of a larger area obtained with the 
NANTEN telescope with  an angular resolution of 2\farcm 7, H{\sc i} 21-cm 
line data taken from the ATCA 
survey, IRAS HIRES data, and radio  continuum data at 4.85 GHz 
from the Parkes survey.
} 
   {We report the detection of molecular gas having velocities in the range 
--56 to --33 \kms\ associated with the western region of RCW\,78.  A few 
patches of molecular gas possibly linked to the eastern faint section are 
detected. The CO emission appears concentrated in a region of 
23\arcmin $\times$18\arcmin\ in size, with a total molecular mass of 
(1.3$\pm$0.5)$\times$10$^5$ \msun, mainly connected to the western section.
The analysis of the neutral atomic gas distribution reveals the \hi\ envelope 
of the molecular cloud, while the radio continuum emission shows a ring-like 
structure, which is the radio counterpart of the 
optical  nebula. 
The gas distribution is compatible with the western section of RCW\,78 having 
originated in the photodissociation and ionization of the molecular gas by 
the UV photons of the WN7 star HD\,117688, and with the action of the stellar
winds of the WR star on the surrounding gas. In this scenario, the 
interstellar 
bubble expanded  more easily  towards the E than towards the W due to
the lack of dense molecular gas in the eastern section. The proposed 
scenario also explains the off center location of WR\,55.  A number
of infrared point sources classified as YSO candidates showed that 
stellar formation activity is present in the molecular gas linked to the
nebula. The fact that the expansion of the bubble have triggered 
star formation in this region can not be discarded. 
 }
   {}


\titlerunning{The molecular environment of RCW\,78}

   \maketitle
%

\section{Introduction}

Interstellar bubbles created by the stellar winds of massive stars are
generally detected as thermal radio continuum shells, as cavities and 
expanding shells in the H{\sc i} 21-cm line emission distribution, 
and as  shells in the far infrared (see Cappa 2006 for a summary). 
Radio observations allowed to estimate the electron densities and 
ionized masses for a number of galactic ring nebulae and to identify their 
neutral gas counterparts.

The dense gas linked to these nebulae has been detected through 
molecular line observations.  However, up to present, only a few cases 
have been analyzed. These observations  revealed the existence of large 
amounts of molecular gas  and that photodissociation
regions at the interface between the ionized and molecular gas and shock 
fronts are present in interstellar bubbles (Cappa et al.
2001, Rizzo et al. 2003).  

In the last years,  infrared images obtained from the MSX Galactic
Plane Survey (Price et al. 2001) and GLIMPSE survey (IRAC images, Benjamin 
et al. 2003) allowed to investigate the  distribution of the near- and 
mid-infrared emission associated with interstellar bubbles, confirming
that PDRs  and shock fronts are common phenomena linked to these structures 
(Churchwell et al. 2006,  Cyganowski 
et al. 2008, Watson et al. 2008, Watson et al. 2009).  

Observational and theoretical studies, as well as numerical
simulations, showed that only a small amount of the stellar wind energy 
released to the interstellar medium is converted into kinetic energy of the 
bubbles (see for example Chu et al. 1983,
Oey 1996, Freyer et al. 2003, 2006, Cooper et al. 2004, Cappa 2006). 

An important issue is the fact that the stellar formation process 
may be favoured in the compressed layers around the nebulae (Elmegreen 2000, 
Thompson et al. 2004).  Using infrared point 
source catalogues, signs of stellar formation activity have been found in the 
dense molecular envelopes surrounding some interstellar
bubbles (e.g. Cappa et al. 2005, Zavagno et al. 2007). 

Thus, observations of molecular gas associated with interstellar 
bubbles provide information essential to investigate  the energetics and
stellar formation process in the dense  surrounding shells.

Here we investigate the distribution of the molecular gas associated with the 
optical ring nebula RCW\,78 around WR\,55 based on high angular resolution 
SEST observations, complemented with lower resolution data of the NANTEN 
telescope.
Complementary IR, \hi\ 21-cm line, and radio continuum archival data allow
the analysis of the distribution of the ionized and neutral atomic 
gas, and that of the dust.  
Our aims are to identify and characterize the material linked to the ring 
nebula and to study its kinematics and energetics.

RCW\,78 (= G\,48b) is a ring nebula of about 35\arcmin\ in diameter. 
The DSS-R image  of the whole nebula is shown in Fig. 1, where the
white cross marks the position of the WR star (see also 
figure 5 by Chu et al. 1983). The brightest part 
of RCW\,78 is about 10\arcmin \por 6\arcmin\ in size and offset to the 
northwest of the star, while fainter regions are present to the northeast,
east, and south (Chu \& Treffers 1981, see also Heckathorn et
al. 1982). Chu (1981) classified the optical nebula as R$_a$ because of
its low expansion velocity and the lack of evidence for a shell structure 
in the velocity pattern of the bright regions. 
This last characteristic is explained by Chu \& Treffers as due to
the fact that the stellar winds have existed during a short period of 
time.

   \begin{figure}
   \centering
   \includegraphics[width=8.5cm]{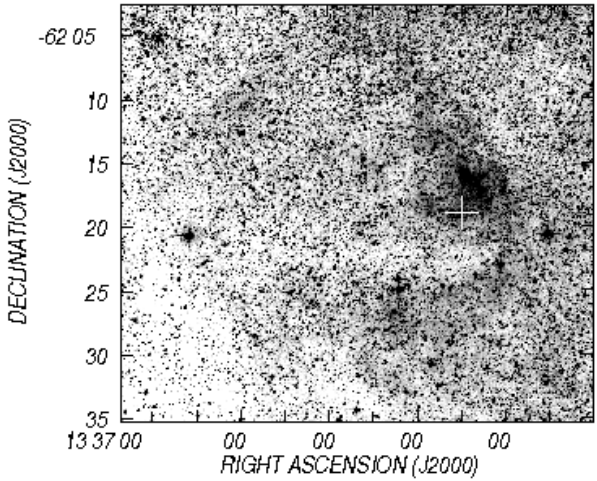}
   \caption{DSS-R image of the eastern and western sections of RCW\,78.
The cross indicates the location of the WR star. }
         \label{FigVibStab}
   \end{figure}

The H$\alpha$ study by Chu \& Treffers (1981) showed that the velocity of the 
ionized  material spans from --38 to --53 \kms,
varying from --44 \kms\ near the star to --53 \kms\ 7\arcmin\ north of it.
Georgelin et al. (1988) found a similar velocity of --41.4 \kms\ based on 
H$\alpha$ Fabry-Perot data. 
Chu \& Treffers (1981) explained the observed 
velocity pattern, which does not correspond to that of an expanding shell, 
as the consequence of an outflow caused by the ionization of the surface 
of the molecular cloud by the central WR star. 
Georgelin et al. (1988) found an additional component in the H$\alpha$ 
line profiles at --24 \kms,  most probably unrelated to the nebula.

   \begin{figure*}
   \centering
   \includegraphics[width=15cm]{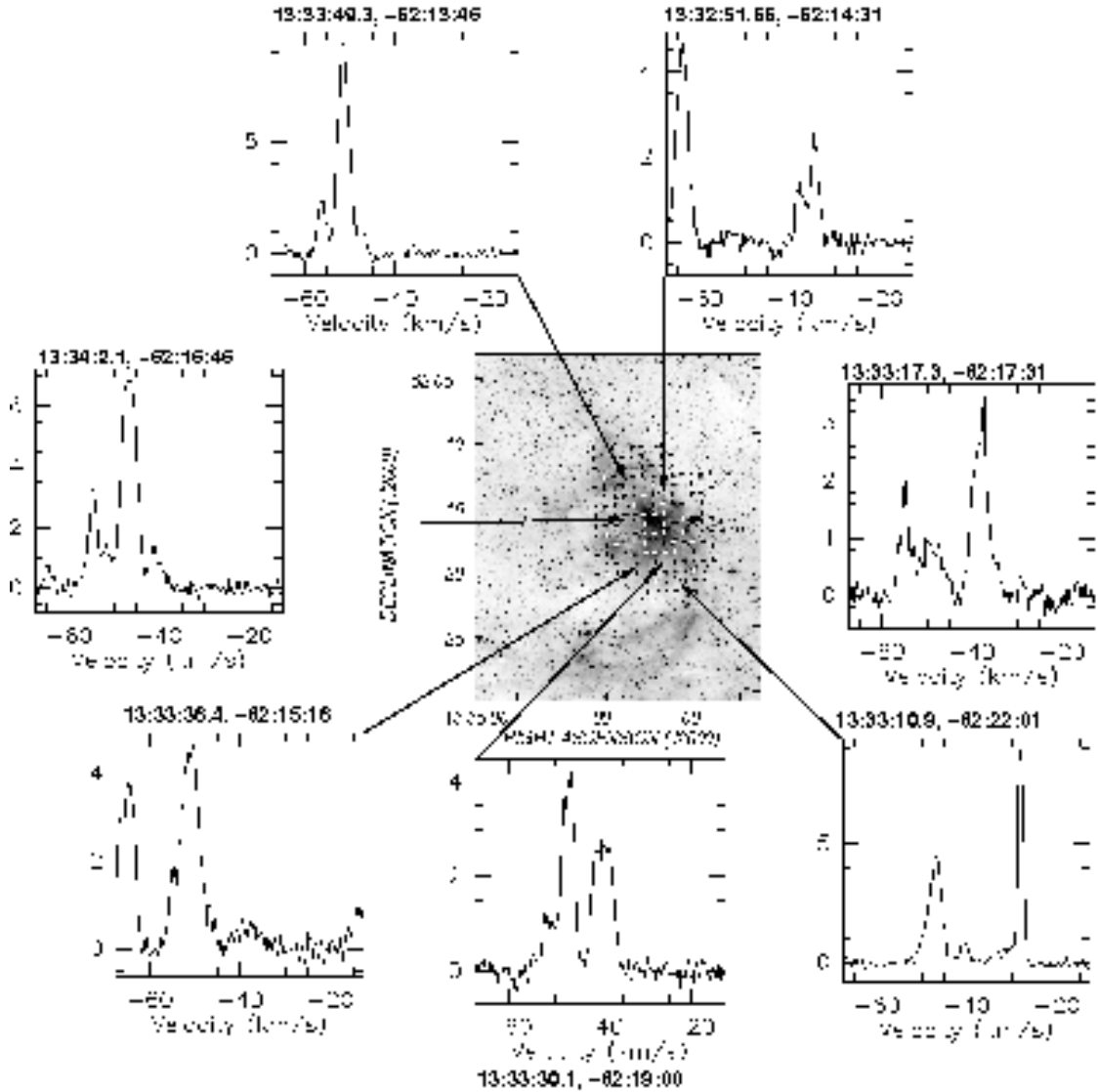}
   \caption{The central panel displays the SuperCOSMOS H$\alpha$ image of 
the W section of RCW\,78.
The grayscale is in arbitrary units. The crosses indicate the position of the 
170 points observed in CO lines. The $^{12}$CO(2-1) spectra corresponding to
selected positions show molecular components with different velocities. The 
position of each spectrum is indicated. Intensities are given in 
main-beam brightness temperature.} 
         \label{FigVibStab}
   \end{figure*}

Circular galactic rotation models (e.g. Brand \& Blitz 1993) predict
that gas with LSR velocities in the range --38 to --53 \kms\ is placed at 
kinematical distances of 3.5-7.0 kpc. In this section of the Galaxy, the 
mentioned velocities are close to that of the tangent point, i.e. --48 \kms. 
 
The study of the ionization structure by Esteban (1993) indicates that 
photoionization is the main source of excitation of the nebula, 
compatible with the clasification by Chu (1981).

The nebula is related to HD\,117688 (= WR\,55 = MR\,49),  a 
WN7 star located at {\itshape (l,b)} = (307\deg 48\arcmin,+0\deg 9\farcm 6)
or \radec\ = (13$^h$33$^m$30.1$^s$, --62\deg 19\arcmin 
1.2\arcsec). Spectrophotometric distances $d$ were estimated by several 
authors:
4.0 kpc (Georgelin et al. 1988), 5.5 kpc (Conti \& Vacca 1990), 6.0 kpc (van 
der Hucht 2001). New NIR calibrations of absolute magnitudes in the  $K_s$ 
band by Crowther et al. (2006), indicate an absolute
magnitude $M_{Ks}$ = --5.92 mag for WN7-9 weak lined stars. Taking into
account the  $K_s$-value for WR\,55 from the 2MASS catalogue (Cutri et al.
2003), and interstellar
extinction values  from Marshall et al. (2006), a distance in the range 
4.5-5.0 kpc can be derived.  
Based on the available distance estimates, we will adopt a distance of 
5.0$\pm$1.0 kpc for the WR star and its surrounding ring nebula.

The terminal wind velocity for WR\,55 derived from the P Cyg profile of 
the CIV $\lambda$1550 line is in the range $V_w$ = 1000 -- 1200 \kms\ 
(Hamann et al. 1993, Rochowicz \& Niedzielski 1995, Niedzielski \& 
Sk\'orzy\'nski 2002). As regards the mass loss rate,  
Hamann et al. (1993) estimate $log$ \.M  = --4.2 \msunyr. 

In this paper, we present high angular resolution SEST observations performed 
in the $^{12}$CO(1-0) and $^{12}$CO(2-1)  lines 
towards the brightest section of the nebula, complemented with lower
angular resolution NANTEN data, \hi\ 21-cm line data, radio continuum 
images, and infrared data of the whole region. In the next sections
we describe the SEST data and analyze the distribution of the molecular,
ionized, and neutral atomic emissions.


\section{Data bases}

\subsection{CO data: observations and data reduction}

The high resolution $^{12}$CO(1-0) (115 GHz) and 
$^{12}$CO(2-1) (230 GHz) data were obtained  during two
observing runs in 13-15 February 2002 and 20-21 March 2003,  
with the 15-m Swedish-European Submillimetre Telescope (SEST)  
at La Silla, Chile. 
The half-power beam-width of the telescope was 44\arcsec\  and 22\arcsec\
at 115 and 230 GHz, respectively. The data were acquired with the high 
resolution acousto-optical spectrometer, consisting of  1000 channels,
with a total bandwidth of  100 MHz and a resolution of 40 KHz, 
corresponding to velocity resolutions of $\simeq$ 0.105 \kms\ at 115 GHz
[CO(1-0) line] and $\simeq$ 0.052 \kms\ at 230 GHz [CO(2-1) line].
Calibration was done using the standard chopper technique. 
The system temperatures were $\approx$ 400 K at 230 GHz and 
$\approx$ 320 K at 115 GHz. 
Pointing was checked once during each observing run on the SiO
(v=1, $J$=2$\rightarrow$1) maser source Ori A. Pointing errors were  
3\arcsec. The uncertainty in the intensity calibration was 10\%.
Details about the telescope and receivers can be found in  
Booth \etal\ (1989).
The observed velocity intervals ranged from --70 to --10 \kms\ at
230 GHz and from --90 to +10 \kms\ at 115 GHz.

  \begin{figure*}
   \centering
   \includegraphics[width=15cm]{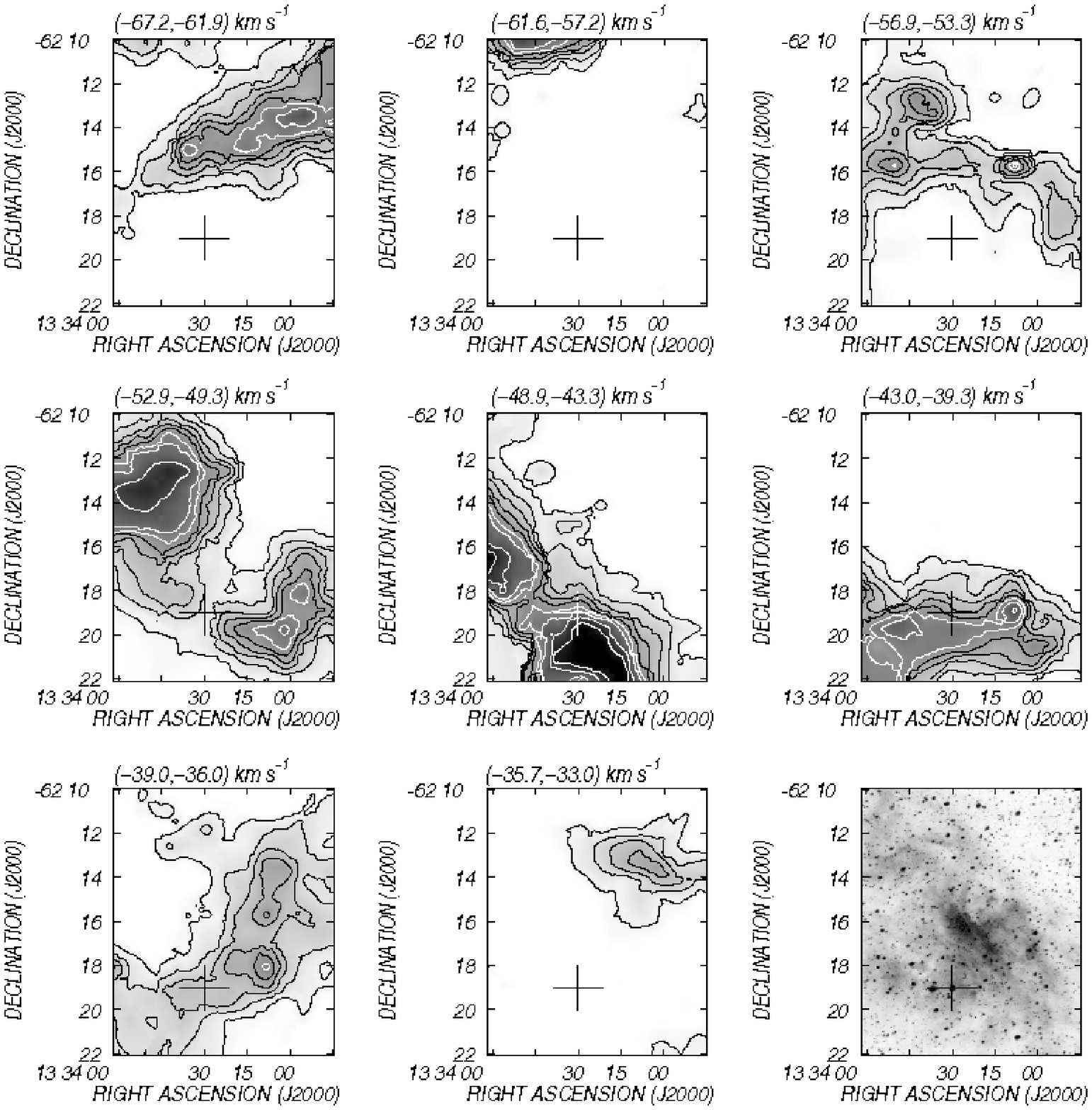}
\caption{Images showing the integrated emission of the CO(2-1) line within
the velocity range from --67 to --33 \kms\ within selected velocity intervals. 
The grayscale is from 1 to 25 K \kms, and the contours are from 2.0 to 14.0
K \kms\ in steps of 2.0 K \kms, and 18.0 and 22.0 K \kms. 
The bottom right  panel displays the SuperCOSMOS image for comparison. 
The cross marks the position of WR\,55.}
   \label{FigVibStab}
   \end{figure*}

$^{12}$CO(2-1) and $^{12}$CO(1-0) lines were acquired simultaneously in the
position-switching mode on a grid with a spacing of 45\arcsec. 
The off-source position, at which no CO emission was detected,  was  
placed at \radec\ = (13$^h$33$^m$10.3$^s$, \hbox{--62\deg 2\arcmin} 
41\arcsec). 
The $^{12}$CO observations were taken in the direction of 170 points towards
the brighest section of the nebula and its environs. The observed positions
are indicated by  crosses in the SuperCOSMOS H$\alpha$ image (Fig. 2).

Aditionally, simultaneous $^{13}$CO(2-1) and $^{13}$CO(1-0) 
observations were carried out 
towards \radec\ = (13$^h$33$^m$10.7$^s$, --62\deg 16\arcmin 1\arcsec) and 
\radec\ = (13$^h$33$^m$10.91$^s$, --62\deg 13\arcmin 46\arcsec),
two positions with strong  $^{12}$CO.
The $^{13}$CO data were acquired using the same 1000 channel high resolution 
spectrometer. The velocity resolution was 0.116 \kms\ at 110 GHz and 
0.058 \kms\ at 220GHz.

The spectra were reduced using the CLASS software (GILDAS working 
group)\footnote{htto://www.iram.fr/IRAMFR/PDB/class/class.html}.  
A linear baseline fitting was applied to the data, except in a few profiles 
where an order 3 polinomial was used.
After smoothing the $^{12}$CO profiles to a velocity resolution of 0.44 
and 0.33 \kms\ at 115 and 230 GHz, respectively, the typical rms noise 
temperature  was  0.22 K ($T_{mb}$) at 230 GHz and 0.11 K  
at 115 GHz after an integration time of 3 min. 
The rms noise temperature of the  $^{13}$CO(2-1) spectra was 0.20 K and
0.10 K for $^{13}$CO(1-0) after an integration time of 6 min.
The spectra were  smoothed to a velocity resolution of 0.46 \kms\ and 
0.35 \kms\ at 110 GHz and 220 GHz, respectively. 

The observed line intensities are expressed as main-beam brightness 
temperatures $T_{mb}$, by dividing the antenna temperature $T^{\ast}_A$  
by the main-beam efficiency $\eta_{mb}$, equal to 0.72 and 0.57
at 115 and 230 GHz, respectively (Johansson et al. 1998). 

$^{12}$CO(2-1) and $^{12}$CO(1-0) data cubes were constructed within
AIPS software, which was used to perform most of the analysis.

\subsection{Complementary data}

Intermediate angular resolution $^{12}$CO(1-0) data obtained with the 4-m
{\rm NANTEN} millimeter-wave telescope of Nagoya University corresponding to 
a region of 1\fdg 5 in size centered at  \radec\ = (13$^h$35$^m$, 
--62\deg 15\arcmin) were used to investigate the large scale 
distribution of the molecular gas in the environs of the whole RCW\,78 
nebula. The half-power beamwidth was  2\farcm 6.  The 4K cooled SIS mixer receiver
provided typical system temperatures of $\approx$ 220K (SSB) at this frecuency. 
The spectrometer was an acoustoptical spectrometer (AOS) with a velocity range of 
100 \kms\ and a velocity reesolution of 0.1 km/s.

The ionized gas distribution was analyzed using data at   
4.85 GHz, extracted from the Parkes-MIT-NRAO (PMN) Southern Radio Survey 
(Condon et al. 1993). This survey was obtained with an angular resolution of
4\farcm 2 and an rms noise of 10 \mjyb. 

To investigate the neutral atomic gas distribution we  extracted \hi\ data 
from the Southern Galactic Plane Survey (SGPS) 
obtained with the Australia Telescope Compact Array (ATCA) and
the Parkes radiotelescope. A Hanning smoothing was applied to these data 
to improve the signal to  noise ratio. The final data cube has a 
synthesized beam of 2\farcm 4$\times$2\farcm 1, a velocity resolution 
of 1.64 \kms, and an rms noise of 1.0 K. A description of this survey 
can be found in McClure-Griffiths et al. (2005).

The distribution of the IR emission was analyzed using high-resolution
(HIRES) IRAS and MSX data obtained through {\it IPAC}\footnote{{\it IPAC} 
is funded by NASA as part of the {\it IRAS} extended mission under 
contract to Jet Propulsion Laboratory (JPL) and California Institute of 
Technology (Caltech).}. The IR data in the {\it IRAS} bands at 60 and 
100 $\mu$m have angular resolutions of 1\farcm 1 and 1\farcm 9, respectively. 
The images in the MSX bands  centered at 8.3, 12.1, 14.7, and 21.3 $\mu$m 
have an angular resolution of 18\farcs 3. 
Images from the Spitzer mid-infrared data at 3.6, 4.5, 5.8, and 8.0 $\mu$m
were obtained from the Galactic Legacy Infrared Mid-Plane Survey 
Extraordinaire (GLIMPSE, Benjamin et al. 2003) and retrieved from the Spitzer 
Science Center\footnote{http://scs.spitzer.caltech.edu}. 
 
  \begin{figure}
   \centering
   \includegraphics[width=7cm]{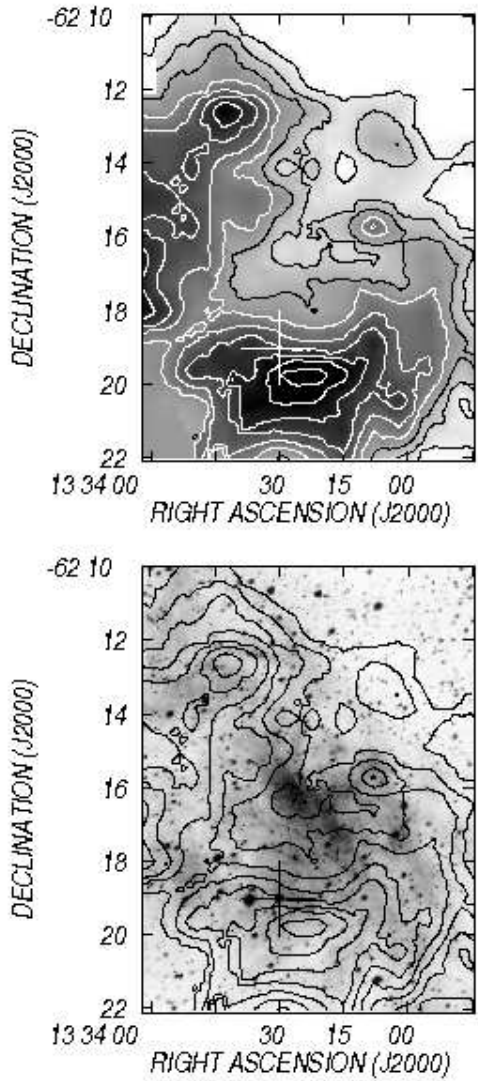}
\caption{Top panel: Image showing the integrated emission of the 
\hbox{CO(2-1)} 
line within the velocity range from --56 to --33 \kms. The 
grayscale is from 8 to 50 K \kms, and the contours are from 10.0 K \kms\
to 50.0 K \kms\ in steps of 5.0 K \kms. Bottom panel: Overlay of the CO 
contours of the top panel and the Super COSMOS image. The cross marks the 
position of WR\,55.}
   \label{FigVibStab}
   \end{figure}

Finally, to search for stellar formation activity in the surroundings of WR55, 
we have inspected available IR point source catalogues: the MSX6C 
Infrared Point Source Catalogue (Egan et al. 1999) in Bands  A 
(\hbox{8.3 $\,\mu$m} ), C (\hbox{12.1 $\,\mu$m}), D (\hbox{14.7$\,\mu$m}), 
and E (\hbox{21.3$\,\mu$m}); the IRAS Point Source Catalogue \footnote{1986 
IRAS catalogue of Point Sources, Version 2.0 (II/125)} at 12, 25, 60, and 
100$\,\mu$m; and the IRAC point source catalogue, which includes photometry 
at 3.6, 4.5, 5.8, and 8.0 $\mu$m. 

\section{The distribution of the molecular gas}

\subsection{Analysis of the SEST data}

Figure 2 displays CO(2-1) profiles obtained towards different 
sections of the nebula to illustrate the molecular components detected towards
RCW\,78. Within the observed velocity range, the CO profiles show molecular 
gas within the velocity interval --67 to --20 \kms. 
The main $^{12}$CO components were detected
at about --65, --50, and --40 \kms. 
Minor components detected in selected areas have velocities of around --55,
--46, --35, and --28 \kms.

Figure 3 displays a series of images spanning the velocity range --67 to  --33
\kms\ showing the distribution of the integrated emission of the CO(2-1) 
line within selected velocity intervals.
The velocity intervals, which are indicated in the upper part of each image, 
were selected to emphasize the different CO velocity components.
The bottom right panel shows the optical H$\alpha$ image 
of the brightest section of the nebula. Components present at different 
velocities are described in the following paragraphs.

The CO emission distribution at velocities in the range \hbox{--67.2} to 
--61.9 \kms\ shows an  elongated cloud extending from E to W, beyond the
surveyed area. Within the velocity interval from --61.6 to --57.2 \kms, the
emission is concentrated towards the higher declination section of the image.

Molecular gas within the range --56.9 to --53.3 presents three
intense  clumps at \radec\ =  (13$^h$33$^m$8$^s$, --62\deg 15\arcmin 45\arcsec),
\radec\ =  (13$^h$33$^m$42$^s$, --62\deg 13\arcmin 45\arcsec), and 
\radec\ =  (13$^h$33$^m$48$^s$, --62\deg 15\arcmin 45\arcsec).   A 
comparison with the optical image shows that the westernmost CO clump is 
located at the same
position of a faint circular optical rim, suggesting that the borders of the 
CO clump  are dissociated/ionized by the strong UV photon flux of the
WR star. The CO emission is weak near the northern extreme of the brightest 
optical filament, whose W border is partially delineated by  CO emission.
   
Gas in the interval from --52.9 to --49.3 \kms\ surrounds the brightest
optical region all around but towards the NW. The CO emission
is strong to the NE and SW. 

Emission in the range --48.9 to --43.3 \kms\ 
surrounds the SE border of the brightest optical filament. 

In the range --43.0 to --39.3 \kms, the emission is parallel to 
{\it Decl.(J2000)} = --62\deg 20\arcmin. This CO cloud coincides with a 
dust lane clearly seen in the H$\alpha$ image. 
Note that the intense CO clump at \radec\ =  (13$^h$33$^m$8$^s$, 
--62\deg 19\arcmin) is projected right to the southwest border of the 
brightest optical filament. 

The emission in the range --39.0 to --36.0 \kms\ partially borders the 
W section of the brightest part of the nebula. The molecular emission at these
velocities is also coincident with regions of strong absorption.  

Finally, the image corresponding to the interval --35.7 to --33.0  
\kms\ displays a cloudlet also coincident with the region of optical absorption
to the northwest of the nebula. 

The CO emission distribution in the whole velocity range and its
comparison with the ionized gas indicates that the molecular material 
detected from --57 to --33 \kms\ is very probably associated with the nebula. 
We believe that CO emission present in the velocity range --67.2 to --61.9 
\kms, which extends beyond the surveyed area and appears well separated in 
velocity from gas at v $>$ --57 \kms, is unrelated to the nebula.  
Fig. 4 shows the CO(2-1) 
integrated emission within the velocity range --57 to --33 \kms\ in
grayscale and contours and an overlay with the SuperCOSMOS image for
comparison. Clearly, the molecular gas emission is enhanced in the environs of 
the brightest region of RCW\,78, encircling it all around. The comparison of
the CO and optical emission distributions indicates that the nebula is 
interacting with molecular gas. 
 
An inspection of table 2 by Chu \& Treffers (1981) shows that the velocities 
of the molecular gas are similar to those of the ionized gas: larger negative 
CO and H$\alpha$ velocities are present to the N of the WR star, while
lower negative CO and H$\alpha$ velocities are detected close and S of the
star. The similar CO and \hii\ velocities reinforces the association of
the ionized and molecular gas.

  \begin{figure*}
   \centering
   \includegraphics[width=13cm]{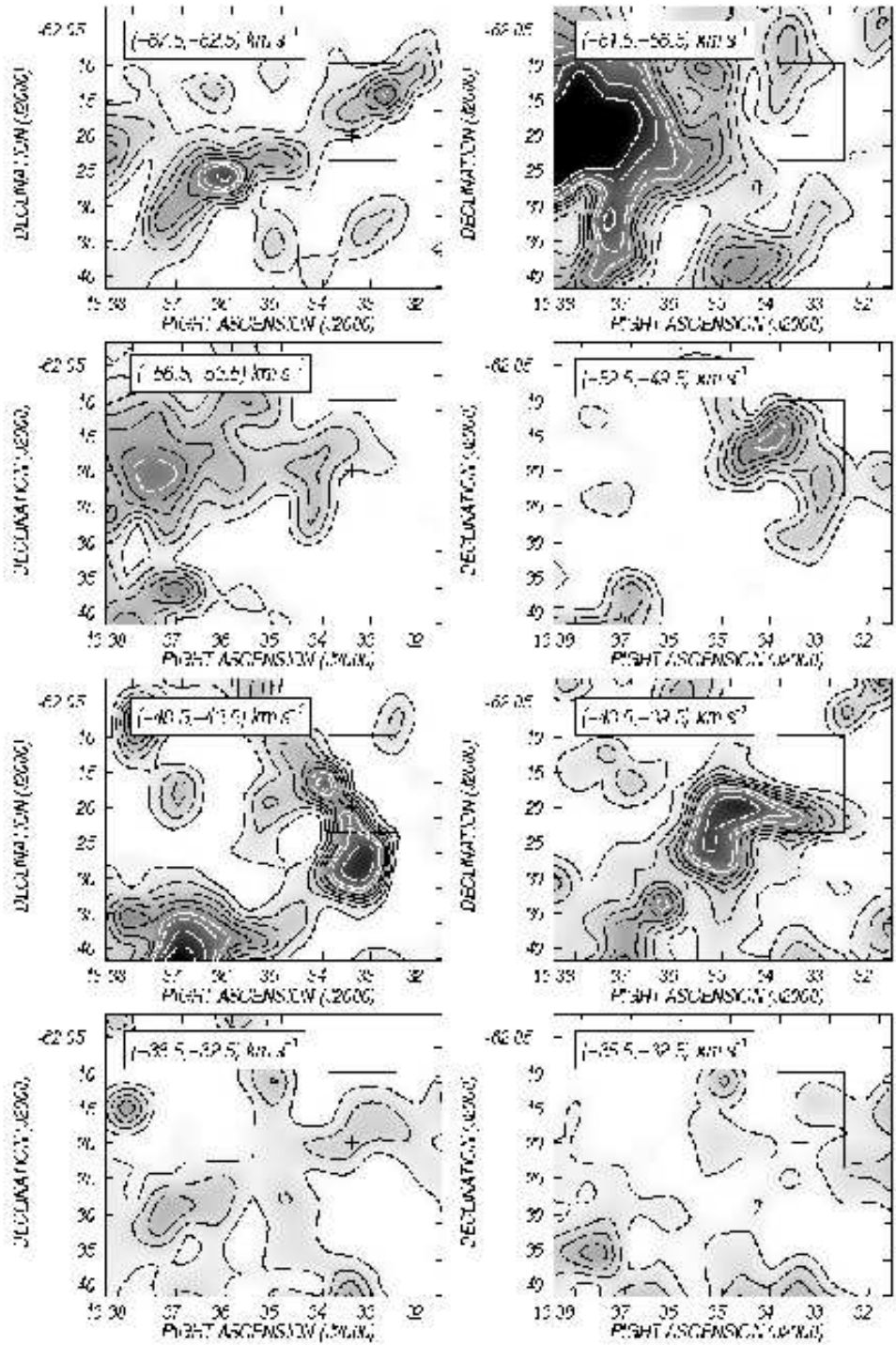}
\caption{Images showing the integrated emission of the CO(1-0) line within
the velocity range from --67 to --33 \kms\ within selected velocity intervals,
as obtained using the NANTEN telescope. 
The grayscale is from 1 to 25 K \kms, and the contours are from 2.0 to 14.0
K \kms\ in steps of 2.0 K \kms, and 18.0 and 22.0 K \kms. The  rectangle
indicates the region observed with the SEST, and the cross marks the 
position of WR\,55.}
   \label{FigVibStab}
   \end{figure*}

\subsection{Analysis of the  NANTEN data}

Figure 5 displays the distribution of the molecular gas in a larger area 
using data from NANTEN. The velocity range of each image has been selected to
facilitate the comparison with the SEST images (Fig. 3). The  rectangle
indicates the area observed with SEST. In the following paragraphs we
describe the NANTEN images, focussing first on the area observed with the 
SEST telescope, and then, extending the analysis to a larger area.

The image corresponding to the velocity range --67.5 to \hbox{--62.5} \kms\ 
shows an elongated CO cloud partially coincident with the brightest part 
of the nebula.
The emission distribution indicates that this cloud is an extension of a 
larger one, which is present to the southeast of the WR star.

  \begin{figure*}
   \centering
   \includegraphics[width=15cm]{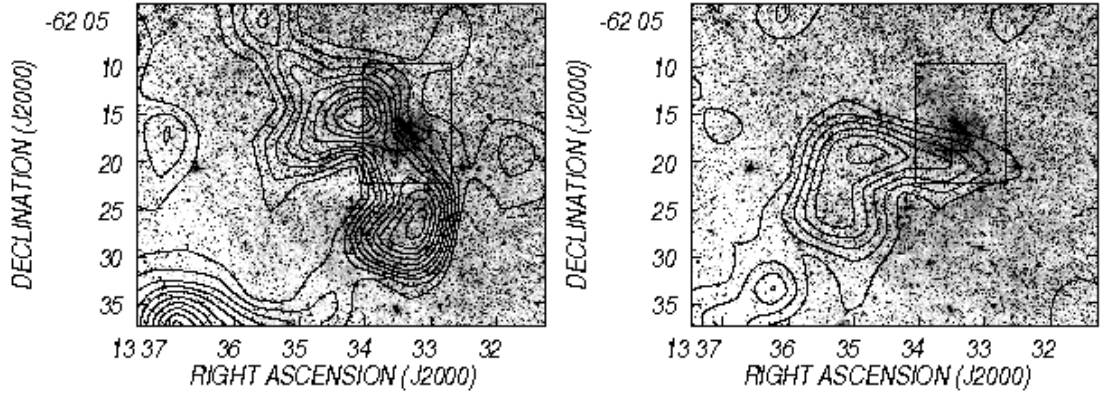}
\caption{{\it Left panel:} Overlay of the DSS-R image and the integrated 
emission of the CO(1-0) line within the velocity range from --56.5 to --43.5 
\kms\ as obtained using the NANTEN telescope. The contours are from 4.0 to 16.0
K \kms\ in steps of 3.0 K, and from 20.0 to 40.0 K \kms\ in steps of 4.0 
K \kms.
{\it Right panel:} Overlay of the DSS-R image and the integrated 
emission of the CO(1-0) line within the velocity range from --43.5 to --39.5 
\kms\ as obtained using the NANTEN telescope. The contours are from 4.0 to 16.0
K \kms\ in steps of 3.0 K \kms, and from 20.0 to 40.0 K \kms\ in steps of 
4.0 K \kms. 
The rectangle and the cross have the same meaning as in Fig. 5.
}
   \label{FigVibStab}
   \end{figure*}

The emission in the intervals --61.5 to --56.5 \kms\ and \hbox{--56.5}
to --53.5 \kms\ shows little correlation with the optical emission. 

The molecular emission in the range --52.5 to --49.5 \kms\ shows 
a CO arc-like structure which is almost coincident with the optical emission 
at $R.A.(J2000) \leq $   13$^h$35$^m$. The intense CO clump centered at 
\radec\ =  (13$^h$34$^m$, --62\deg 15\arcmin) delineates  the NE section of 
the brightest part of the nebula, while a fainter clump is detected to the SW
(\radec\ =  (13$^h$33$^m$10$^s$, --62\deg 21\arcmin). Both clumps, as well as
the fainter CO emission region in between are easily identified in the SEST
images. This material is very probably related to RCW\,78.

The emission within the range --48.5, --43.5 \kms\ displays a 
CO feature running from \radec\ =  (13$^h$35$^m$, --62\deg 10\arcmin) to 
(13$^h$33$^m$, --62\deg 32\arcmin). The emission coincides with bright 
portions of the nebula and is shifted slightly towards the SW in comparison 
with the emission in the previous image.  This characteristic is
also observed in the image corresponding to the SEST data.

An elongated CO structure having velocities in the range --43.5 to 
--39.5 \kms, present at \radec\ =  (13$^h$34$^m$, --62\deg 20\arcmin) 
appears projected onto a dust lane present at Decl. =  --62\deg 22\arcmin. 
The molecular emission extends towards the south at R.A.(J2000) =  
13$^h$35$^m$.

The CO emission is weak for velocities more positive than --38.5 \kms. 
Within the range --38.5 to --35.5 \kms, a CO cloud is projected onto the 
nebula. The emission distribution changes for v $>$ --35.5 \kms, where the 
CO emission coincides with regions showing faint optical emission.

In summary, the bulk of the molecular emission is concentrated in  two
structures having velocities in the range --52.5 to --43.5 \kms\ and
--43.5  to --39.5 \kms, coincident with the velocities of the 
H$\alpha$ line (Chu \& Treffers 1981). Overlays of both features with the 
optical emission are presented in Fig. 6.

Based on morphological and kinematical evidences we can conclude that the 
arc-like structure having velocities in the range --52.5 to --43.5 \kms\ is 
clearly associated with RCW\,78 (Fig. 6, left panel). 
As regards the feature with velocities from --43.5 to --39.5 \kms\ (Fig. 6, 
right panel), its velocity is also coincident with those of the ionized gas, 
suggesting a relation to the nebula. 
On the contrary, material having velocities lower than --57 \kms\ does
not seem to be related to the nebula.

CO emission linked to the  eastern section of the nebula (near 13$^h$36$^m$) 
is difficult to identify.
Molecular gas  related to the eastern section is probably present in the 
range --43.5 to --38.5 \kms\ at  \radec\ =  (13$^h$36$^m$, --62\deg 5\arcmin) 
and at  \radec\ =  (13$^h$37$^m$, --62\deg 17\arcmin). However, the CO emission
is weak, suggesting that the amount of molecular gas associated with 
this region is small as compared with the western regions.

Both the SEST and NANTEN data in the environs of the western 
section of RCW\,78 are consistent with a scenario in which the ionized gas 
originated through photodissociation and ionization of the parental 
molecular cloud. The lack of dense gas towards the east  has probably 
favored the expansion of the ionized gas in this direction. 

Adopting a mean radial velocity of around --45 \kms, circular galactic 
rotation models (e.g. Brand \& Blitz 1993) predict near and far kinematical 
distances of 4.2 and 6.4 kpc, respectively, in agreement with the 
spectrophotometric distance of WR\,55.

\section{The associated ionized  and neutral atomic gas}

\subsection{Ionized gas distribution}

The upper panel of Fig. 7 displays the radio continuum image at 4.85 GHz in 
contours and grayscale. The location of the WR star is marked by 
the cross. The overlay of 
the optical and radio images shows that RCW\,78 is detected at 
this frequency. The brightest section of the nebula coincides with one
of the most intense radio continuum regions.

The brightest part of RCW\,78 at \radec\ = (13$^h$33$^m$30$^s$, 
--62\deg 17\arcmin) coincides with the brightest radio emission region. 
The comparison with the CO emission distribution shows that the intense radio 
emitting region is interacting with molecular material.

  \begin{figure}
   \centering
   \includegraphics[width=8cm]{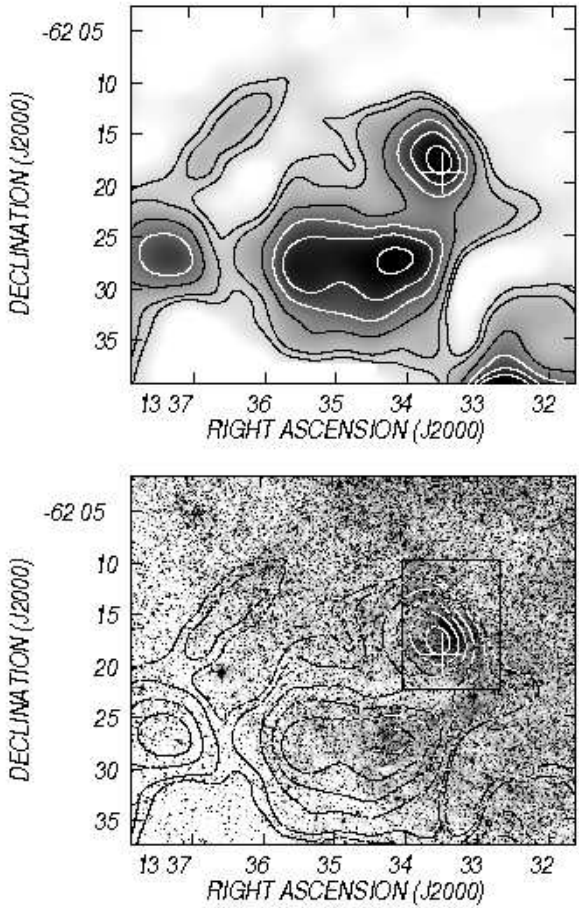}
\caption{{\it Top panel:} Radio continuum image at 4.85 GHz. The grayscale is
from --0.05 to 0.25 \jyb, and the contour lines are 30, 50, 100, 150, 200, 
and 250 \mjyb. 
{\it Bottom panel:} Overlay of the radio continuum image in contour 
lines and the DSS-R image. The rectangle delineates the region observed with
SEST. }
   \label{FigVibStab}
   \end{figure}

The radio continuum emission areas at \radec\ =  (13$^h$34$^m$50$^s$, 
--62\deg 27\arcmin) and \radec\ =  (13$^h$36$^m$30$^s$, --62\deg 15\arcmin) 
coincide with fainter optical emission regions. The region at 
\radec\ =  (13$^h$33$^m$40$^s$, --62\deg 22\arcmin), which lacks strong radio 
continuum emission, coincides with molecular gas having velocities
in the range --43.5 to --39.5 \kms \ (see Figs. 2 and 4). 

The strong radio emission region at \radec\ = (13$^h$37$^m$15$^s$, 
--62\deg 27\arcmin), which is projected onto a region of strong absorption 
bordering faint optical areas belonging to RCW\,78, is very probably 
unrelated.

The ring appearance of the radio continuum emission, which resembles the
H$\alpha$ emission distribution, is suggestive of the action of the stellar
winds of massive stars on the surrounding gas, which shaped an interstellar 
bubble. The ionized gas  has probably expanded more easily  towards the 
E than towards the W due to
the lack of dense gas molecular gas in the eastern section. The proposed 
scenario also explains the off center location of WR\,55.

\subsection{\hi\ gas distribution}

The \hi\ gas emission  distribution in the range 
--55 to --25  \kms\ is complex and  clumpy, with few clear structures. 
Figure 8 displays the \hi\ emission distribution  within the velocity interval 
from --50 to --40 \kms\ towards RCW\,78. 

The region of low \hi\ emission centered at
\radec\ =  (13$^h$33$^m$50$^s$, --62\deg 23\arcmin) coincides with the 
western section of the nebula (shown in  Fig. 1) and with CO emission
having velocities in the range --52.5 to --39.5 \kms\ (see Fig. 5), suggesting
that most of the neutral gas in this region is H$_2$.  In this scenario,
part of the neutral atomic gas encircling the depression might correspond 
to the envelope of the molecular cloud. 

  \begin{figure}
   \centering
   \includegraphics[width=8.5cm]{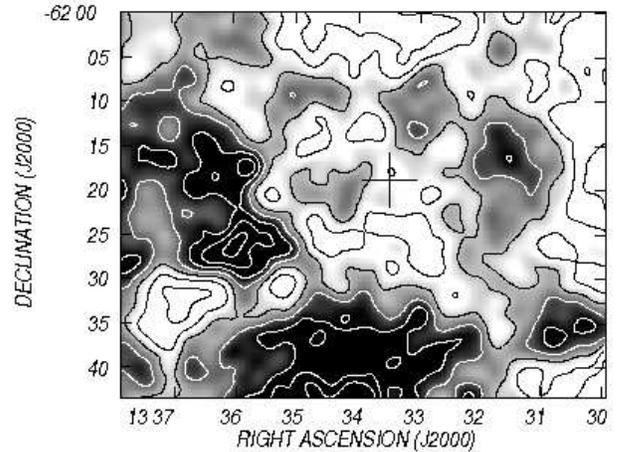}
\caption{Mean \hi\ brightness temperature distribution within the velocity
interval from --50.0 to -40.0 \kms. The grayscale is from 87 to 110 K, and
the contour lines are from 80 to 110 K in steps of 5 K. The cross marks 
the position of WR\,55.}
   \label{FigVibStab}
   \end{figure}

\subsection{Far IR emission}

The HIRES IRAS image at 60 $\mu$m is shown in the upper panel of 
Fig. 9. The source at  
\radec\ =  (13$^h$33$^m$20$^s$, --62\deg 17\arcmin) coincides with the 
brightest section of RCW\,78, indicating the presence of dust mixed with 
the ionized gas. We will refer to this source as source A.

The region at Decl. =  --62\deg 26\arcmin, from  R.A. =  13$^h$32$^m$30$^s$ 
to 13$^h$36$^m$, which depicts relatively strong far IR emission, coincides 
with an area showing optical emission. Two bright extended IR sources can be 
identified: one at \radec\ =  (13$^h$34$^m$15$^s$, --62\deg 26\arcmin) 
(from here on referred  to as source B), and 
the other at \radec\ =  (13$^h$35$^m$10$^s$, --62\deg 26\arcmin) (source C).
The sparcely populated open cluster C1331-622 and the 
star HD\,117797 (\radec\ =  13$^h$34$^m$12.0$^s$, 
\hbox{--62\deg 25\arcmin 1\farcs 8,} O8Ib(f), Walborn 1982) are 
projected onto the line of sight to source B.
C1331-622 is an open cluster of A and F-type stars placed at 820 pc, while the
estimated distance for the O-type star is  3.9 kpc (Turner \& Forbes 2005).

Based on the IR fluxes at 60 and 100 $\mu$m, and following the procedure 
described by Cichowolski et al. (2001), we derived the dust color temperature 
for  sources A, B, and C. 
Dust color temperatures are $T_d$ =  29$\pm$7 K, 44$\pm$7 K, and  48$\pm$6 K
for sources A, B, and C, respectively. 
The range of temperatures for each source corresponds to $n$ = 1-2 and to 
different IR background emissions. The parameter $n$ is related to the dust 
absorption efficiency ($\kappa_\nu\ \propto\ \nu^n$). 
The dust temperature derived for the far IR source related to the brightest
region of RCW\,78 (source A) is typical of \hii\ regions. Estimates for sources
B and C are clearly higher, probably revealing the presence of inner 
heating sources.

The IRAC image at 8 $\mu$m displays very faint extended emission 
towards the brightest section of RCW\,78 (coincident with the IR source A), 
without identifiable structures. The emission in this band towards 
IR sources B and C shows some weak small scale features which are 
difficult to characterize without high angular resolution molecular line 
data. The emission in MSX Band E (at 21.3 $\mu$m) shows a region of 
emission of about 2\arcmin\ in 
size, centered at \radec\ =  (13$^h$34$^m$13$^s$, --62\deg 25\arcmin), 
approximately coincident with source B.  
Band E includes continuum emission from very small grains, which 
can survive  inside ionized regions (cf. Deharveng et al. 2005 and 
references there in), and 
a contribution from nebular emission lines. 

\section{Stellar formation}

To look for stellar formation activity in the surroundings of WR\,55, we have 
inspected three different infrared point source catalogues: 
Spitzer, MSX, and IRAS. The study consisted of identifying 
candidates to Young 
Stellar Objects (YSOs)  associated with the molecular gas in RCW\,78.
An area of 30\arcmin $\times$30\arcmin\ in size centered at 
 \radec\ =  (13$^h$34$^m$, --62\deg 20\arcmin)  was examined. 

To identify YSO candidates in the IRAS and MSX catalogues, we followed the 
criteria by Junkes et al. (1992) and Lumsden et al. (2002), 
respectively.

A total number of 20 IRAS point sources were found projected 
onto the analyzed region. Eight out of the 20 sources have IR 
fluxes compatible with protostellar objects, and only 1 of them  is 
projected onto the molecular gas related to the nebula. 

Lumsden et al. (2002)'s conditions were used for MSX sources. The selection of 
candidates was performed taking into account their loci in the 
(F$_{21}$/F$_{8}$, F$_{14}$/F$_{12}$) diagram. 
Their criteria allow identification of candidates to massive young 
stellar objects (MYSOs) and compact \hii\ regions (CHII).  
A total of  16 MSX sources are projected onto the whole region.  
Two sources classified as CHII and 2 MYSO candidates are seen projected 
onto the molecular material associated with RCW\,78. 

  \begin{figure}
   \centering
   \includegraphics[width=8.5cm]{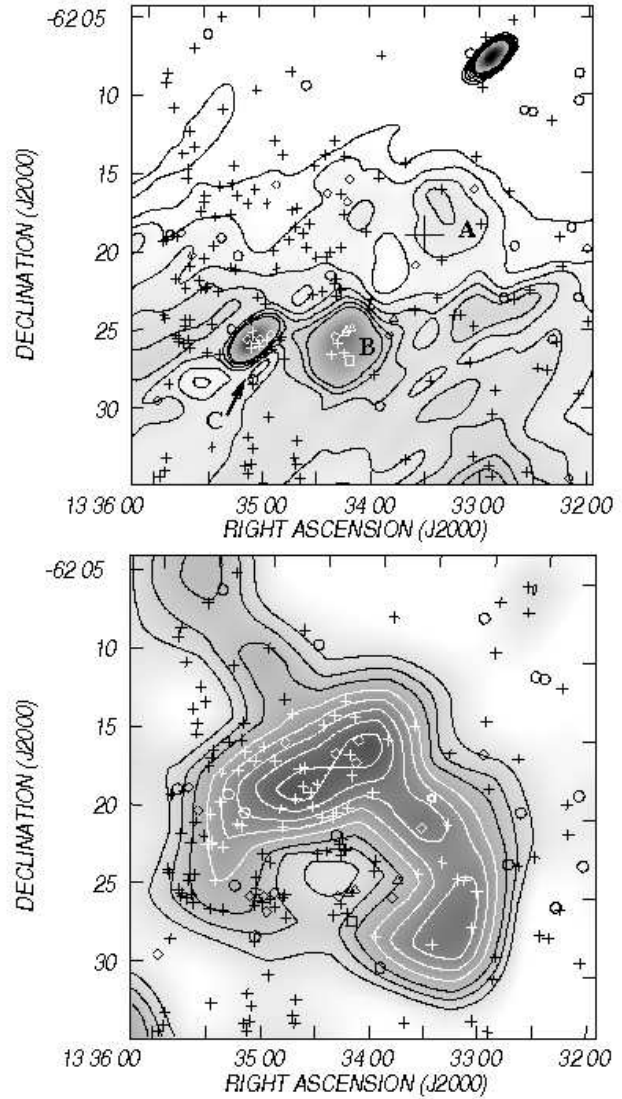}
\caption{  Upper panel: Distribution of the IR emission at 60 $\mu$m. The 
grayscale goes from
170 to 700 MJy ster$^{-1}$, and the contour lines are from 160 to  280 
MJy ster$^{-1}$ in steps of 20 MJy ster$^{-1}$. Letters A, B, and C indicate 
infrared extended sources. 
Bottom panel: CO emission distribution within the velocity range from --56.5 
to --39.5 \kms. The grayscale goes from 4 to 70 K \kms. The contours are
12, 16, 20, 25, 30, 35, 40, and 45 K \kms. The large cross
indicates the position of the WR star. Small crosses, open circles, open 
diamonds, 
open triangles, and the square mark the position of YSO candidates. } 
   \label{FigVibStab}
   \end{figure}

To search for YSO candidates, we also analyzed the mid-infrared emission 
of point sources using Spitzer IRAC photometry. This data base provides 
excellent tools for identifying and classifying YSO candidates as Class I 
and II.
We investigated the infrared excess by measuring the infrared spectral index 
$\alpha_{IR}$ (= $d \log{ (\lambda F_{\lambda})}/d \log{\lambda}$) 
(see Lada 1987, 
Stahler \& Palla 2005, Chavarria et al. 2008, Koenig et al. 2008). 

Within the selected region, we found 5860 sources with detections in the
6 bands ($H$ ans $K_S$, from the 2MASS catalog, and the four IRAC bands).
To deredden the magnitudes, we estimated the extinction using the
2MASS $(H - K_S)$ colors in areas of about 2\arcmin\ in size following 
the method described by Lada et al. (1994). The extinction laws are
from Rieke \& Lebofsky (1985) and Flaherty et al. (2007) for the 2MASS
and IRAC bands, respectively.

To select young candidates, we took into account sources having spectral index
$\alpha_{IR}$ in the range --1.5 to +2 for each pair of IRAC wavelengths. 
Starburst galaxies and active galactic nuclei were removed from the sample
by using the color-color diagram [3.6]-[4.5] vs. [4.5]-[8.0]  from 
Simon et al. (2007). 
We applied a second selection criterium to  the remaining sources, 
taking into account  sources with $(K_s - [4.5]) >$ 
0.4, following Koenig et al. (2008). After inspecting visually the SEDs of the
remaining sources, we were left with 193 YSO candidates.
Note that two selection criteria were applied.
This procedure is a bit strict and we are aware that a number of
YSO candidates have probably been rejected. However, selected sources
are more confident candidates. 

Finally, we used the IR spectral index to discriminate Class I sources 
(0 $< \alpha  <$ +2) from the more evolved Class II sources
(--1.5 $< \alpha <$ 0). Sources with a nearly flat IR spectral distribution
($\alpha\ \simeq$ 0) were named Class I-II. 

The distribution of the YSO candidates identified using the IRAS, MSX, and 
Spitzer data bases is shown in  Fig. 9, overlaid onto 
the 60 $\mu$m and the CO emissions. Note that the CO emission shown in the
figure includes molecular gas at two different velocities (from --56.5 to
--43.5 \kms, and from --43.5 to --39.5 \kms). 
The  symbols used in the figures
have the following meaning: the open square marks the position of the IRAS 
point source, open circles indicate Class I objects; small crosses,  
Class II objects, diamonds Class I-II objects, and open triangles 
correspond to  CHII and MYSOs. 

The spatial distribution of the candidates is not uniform. About 120 sources  
are projected onto the molecular gas, mainly concentrated towards 
{\it R.A.(J2000)} $>$ 13$^h$34$^m$. A relatively large number of 
YSO candidates are projected bordering the low CO emission region 
centered at about \radec\ =  (13$^h$34$^m$30$^s$, --62\deg 25\arcmin).
Two areas in this region are particularly interesting: 
one at \radec\ =  (13$^h$34$^m$15$^s$, --62\deg 25\arcmin) and the other at
\radec\ =  (13$^h$35$^m$, --62\deg 26\arcmin), which include 22 sources.
These sources are listed in Table 1. Their main parameters (coordinates, 
fluxes) as well as an indication of the nature of the sources are included in 
Table 1. 

The upper panel of Fig. 9 shows that both groups of YSO candidates
coincide with the IR extended emission regions named B and C (see Sect. 4.3).  
The MSX sources \#2, \#4, and \#5, the Spitzer sources \#6, \#7, \#8,
and \#9  match the location of the  
IR source B. In particular, MSX sources \#2 and \#4 approximately 
coincide in position with the extended source detected in Band E  and with the 
O8Ib(f) star HD\,117797. 
The presence of this bunch of point sources suggests the existence of 
an active star forming region. The region of extended emission in Band E, 
which is most probably related to a compact \hii\ region in this 
direction, reinforces this suggestion.
Note that the O8Ib(f) star HD\,117797 is projected  onto the 
borders of the bright sections of the  molecular clouds,   
indicating that the 
dense material has probably been dissociated by the UV
photons of this massive star. 

The position of MSX point source \#3 and the Spitzer sources \#10 to 
\#22   coincide with the extended IR source B (at  \radec\ =  
(13$^h$35$^m$15$^s$, --62\deg 26\arcmin), indicating the detection 
of an additional stellar forming region.
The  source IRAS13316-6210 (\radec\ =  (13$^h$35$^m$2$^s$, 
--62\deg 25\arcmin 30\arcsec), whose IRAS fluxes do not match Junkes et al.'s
criteria, is also coincident with this region.

In Figure 10 we plot the dereddened color-magnitude diagram for
the YSO candidates detected using Spitzer database, which appear
related to the IR sources B and C, i.e. those listed in Table 1.
A distance of 5 kpc was assumed. We compare the position of the YSO candidates 
with the main sequence stars from B3V to O3V, and with an isochrone 
calculated by Siess et al. (2000) corresponding to an age of 1 Myr (dot 
line). The K$_s$-limit of our sample is about 14 mag, which corresponds 
to a  limit in the absolute magnitude of K$_S$ of about 0.5 mag and to 
pre-main sequence stars with
masses larger than about 2.2 \msun\ in the Siess et al.'s model. This diagram
suggests that the YSOs candidates in both selected regions have masses 
larger than this last value. Note, however, that as their luminosity 
originates both in the central object and in the circumstellar material,
the effective mass of the candidates may be lower than estimated (Comeron et 
al. 2005).

Sub-millimetre continuum observations would be useful to confirm 
the nature of the YSO candidates. 
   \begin{figure}
   \centering
   \includegraphics[width=8.5cm]{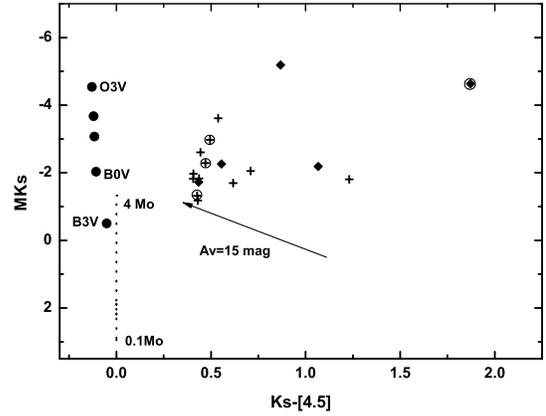}
   \caption{ Color-magnitude diagram for the SPITZER YSO candidates in the
two selected regions near \radec\ =  (13$^h$34$^m$30$^s$, --62\deg 25\arcmin). 
Crosses and diamonds correspond to Class II and
Class I-II objects, respectively. Main sequense stars from O3V to B3V
(dots) and the isochrone for an age of 1 Myr (dotted line) are also plotted. 
The vector represents an absorption of 15 mag. The four sources shown as a 
cross inside a small circle are projected onto IR source B. }
   \end{figure}

\begin{table*}
\centering
\caption[]{YSO candidates projected onto the infrared extended sources B and
C (near  \radec\ =  
[13$^h$34$^m$30$^s$, --62\deg 25\arcmin]). }
\begin{tabular}{rllcrrrrrrl}
\hline
\hline
 $\#$ &    $l$[$h \ \ m \ \  s$]    &  $b$[$\circ\ \arcmin \ \arcsec$]    & {\it IRAS} source  
&     \multicolumn{4}{c}{Fluxes[Jy]}  & &  & Class.\\
 & & &  &  12$\mu$m  &    25$\mu$m &   60$\mu$m &  100$\mu$m &  \\
\hline
1 &  13 34 11.4 & --62 27 03 & 13307-6211 & 0.8  &  2.0  & 20.5 & 85.4   &  &  & YSO/Class 0\\
\hline
\hline
$\#$ &   $l$[$h m s$] &  $b$[$\circ\ \arcmin \ \arcsec$] &   {\it MSX} source 
&     \multicolumn{4}{c}{Fluxes[Jy]} \\     
 &  &  &  &  8$\mu$m &  12$\mu$m & 14$\mu$m & 21$\mu$m &  & &Class. \\
\hline
 2 & 13 34 13.0 & --62 25 07 & G307.8603+00.0439 &  0.2261 &  1.809  & 1.118  &  1.391  & & & C\hii, HD\,117797 \\
 3 & 13 35 03.6 & --62 25 77 & G307.9563+00.0163 &  0.7808 &  0.7287 & 0.5542 &  3.093  & & & C\hii\ \\
 4 & 13 34 10.2 & --62 25 00 & G307.8561+00.0463 &  0.2149 &  1.144  & 2.191  &  4.793  & & & MYSO, HD\,117797 \\
 5 & 13 34 13.8 & --62 25 29 & G307.8620+00.0399 &  0.1614 &  1.146  & 1.726  &  4.996  &  & & MYSO \\
\hline
\hline
$\#$ &  $l$[$h \ \ m \ \ s$]   &  $b$[$\circ \ \arcmin$]  &   {\it GLIMPSE source} & \multicolumn{6}{c}{Fluxes[mag]} \\
 &    &    &   & $H$ & $K_s$ & $[3.6]$ & $[4.5]$ & $[5.8]$ & $[8.0]$ & Class. \\
\hline
6&13 34 14.4384&-62 26.5794&G307.8597+00.0185&13.244&11.95&11.237&11.137&10.873&10.895&II\\
7&13 34 17.9136& -62 25.9506& G307.8680+00.0277& 12.459& 11.257& 10.338& 10.419& 10.242& 10.243& II\\
8&13 34 19.4304&-62 25.506&G307.8721+00.0345&11.202&9.599&7.489&7.385&6.928&6.641&I-II\\
9&13 34 21.564&-62 26.6934&G307.8729+00.0143&13.959&12.913&12.267&12.144&12.116&11.853&II\\
10&13 34 47.6856&-62 26.9472&G307.9219+00.0018&13.37&12.553&11.859&11.738&11.713&11.68&II\\
11&13 34 48.9096& -62 25.4874& G307.9283+00.0254& 14.267& 12.681& 11.796& 11.655& 11.204& 11.654&  II\\
12&13 34 50.1048&-62 25.6944&G307.9300+00.0216&13.767&12.546&11.785&11.7&11.347&11.414&II\\
13&13 34 54.0384&-62 26.2866&G307.9358+00.0106&12.596&11.773&11.018&10.92&10.938&10.933&II\\
14&13 34 54.8448&-62 25.3746&G307.9399+00.0253&13.319&12.117&11.293&11.152&10.792&10.611&I-II\\
15&13 34 56.6544&-62 25.7538&G307.9422+00.0185&14.487&12.589&11.331&10.94&10.732&10.827&II\\
16&13 34 58.644&-62 26.6052&G307.9436+00.0039&13.376&12.672&12.142&11.819&11.508&11.403&I-II\\
17&13 35 0.6672&-62 25.641&G307.9502+00.0190&11.013&9.211&8.089&7.925&7.462&7.366&I-II\\
18&13 35 0.8784& -62 26.061& G307.9494+00.0121& 13.54&	12.424& 11.79& 11.598& 11.264& 11.535& II \\
19&13 35 4.7112&-62 25.1676&G307.9592+00.0255&13.87&12.348&11.382&11.22&10.892&10.994&II\\
20&13 35 5.0928&-62 24.3822&G307.9621+00.0383&14.298&13.186&12.453&12.352&12.217&12.064&II\\
21&13 35 5.5032&-62 26.2302&G307.9577+00.0078&12.143&10.783&9.836&9.826&9.511&9.568&II\\
22&13 35 7.9968&-62 25.6524&G307.9641+00.0165&14.241&12.207&10.879&10.721&10.298&10.135&I-II\\
\hline
\hline
\end{tabular}
\end{table*}
         
\section{Discussion}

\subsection{Parameters of the gas}

The distribution of the molecular gas and its comparison with that of the
ionized gas indicates that the CO emission detected in the range --56 to
--33 is probably related to the nebula. We note that Georgelin et al. (1988) 
proposed 
that the dust lane at Decl. =  --62\deg 22\arcmin\ is unconnected to RCW\,78.
However, the velocity of the molecular emission in this region, which coincides
with the velocity of the ionized gas casts doubts on this interpretation.
Although we can not discard Georgelin et al.'s interpretation, we believe 
that gas in the velocity interval mentioned above is linked to the nebula.

The mean $H_2$ column density $N_{H2}$, the CO luminosity $L_{CO}$,  the
molecular mass $M_{H2}$ and a rough volume density $n$ were derived separately 
for the molecular gas related to RCW\,78 from the SEST and NANTEN data.
The last data give  more reliable values since
the whole molecular cloud is included. 
Mean $H_2$ column densities were derived from 
$^{12}$CO(1-0) data, by making use of the empirical relation between 
the integrated emission $I_{CO}$ 
(= $\int T_{mb} dv)$ and $N_{H2}$. We adopted $N_{H2}$ = 
2.3 $\times\ I_{CO} \times$ 10$^{20}$ cm$^{-2}$ (K \kms)$^{-1}$ 
(Strong et al. 1988). The CO luminosities are defined as 
$L_{CO}$ = $I_{CO} \times A$ (K \kms\ pc$^2$), where $A$ is the area in 
pc$^2$. The mass includes a Helium abundance of 10\%. 

From SEST data we obtained  
$N_{H2}$ = 7.7 $\times$ 10$^{21}$ \cmdos. Adopting $d$ = 5$\pm$1 kpc, 
$L_{CO}$ = (5.9$\pm$2.2)$ \times$ 10$^3$ (K \kms\ pc$^2$) and $M_{H2}$ = 
(3.0$\pm$1.2)\por 10$^{4}$ \msun. 
The volume density $n \simeq$ 750$\pm$130 \cmtres\ was 
estimated by distributing the molecular material within a region of 
5\farcm 1 in radius  (7.4 pc at 5.0 kpc).
The corresponding values derived from the NANTEN data are:
$N_{H2}$ = 6.6 $\times$ 10$^{21}$ \cmdos,  
$L_{CO}$ = (2.7$\pm$1.0)$\times$ 10$^4$ (K \kms\ pc$^2$) and $M_{H2}$ = 
(1.3$\pm$0.5)$\times$ 10$^{5}$ \msun. 
A volume density $n \simeq$ 270$\pm$60 \cmtres\ was estimated 
by distributing the molecular material within a region of 11\farcm 8 in 
radius.

  \begin{figure*}
   \centering
   \includegraphics[width=12cm]{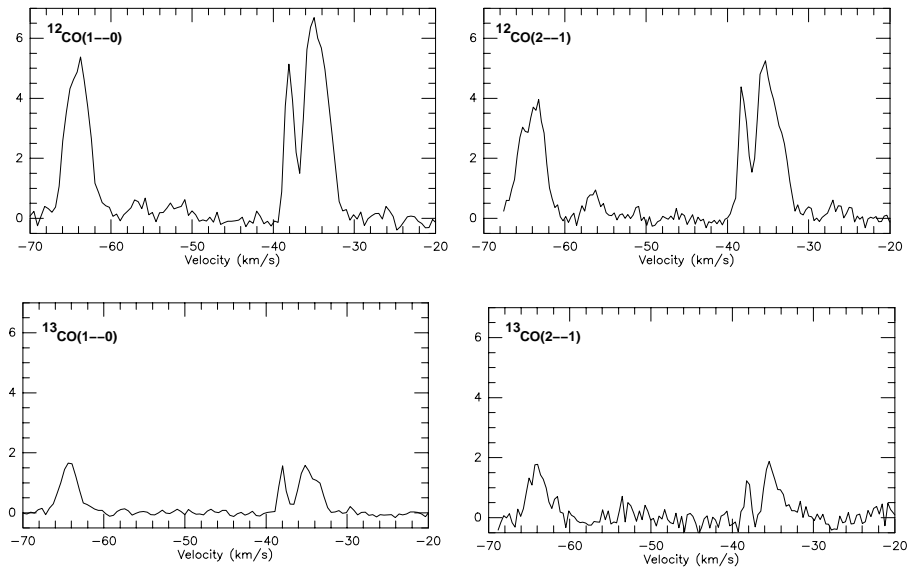}
\caption{$^{12}$CO and $^{13}$CO SEST profiles corresponding to the 
position   \radec\ =  (13$^h$33$^m$10.9$^s$, --62\deg 13\arcmin 46\arcsec). 
Line intensities are expressed as main-beam brightness temperature.}
   \label{FigVibStab}
   \end{figure*}

Optical depths were derived for two positions for which $^{13}$CO profiles
were obtained, assuming that the emission fills completely the SEST beam. 
The $^{12}$CO and $^{13}$CO profiles corresponding to one of these positions 
are shown in Fig. 10. The parameters of the gaussian fitting to the
CO components (peak temperatures  T$_{p}$, central velocities V$_{CO}$,
velocity width $\Delta$V), and the integrated emission $I_{CO}$ are 
listed in Table 2. 

The peak optical depth of the $^{13}$CO(2-1) lines were estimated from the
peak $T_{mb}$ values as $\tau$($^{13}$CO[2-1]) $\simeq -ln\{1 - 
T_{mb}(^{13}$CO[2-1])/[10.58 $J(T_{exc})$ -- 0.21]\}, where $J(T_{exc}) =
[e^{10.58/Texc} - 1]^{-1}$. Excitation temperatures
$T_{exc}$ were derived from the $^{12}$CO(2-1) line assuming LTE and
that $^{12}$CO(2-1) is optically thick. Peak optical depths of the
$^{12}$CO lines  were obtained as  $\tau$($^{12}$CO) = 
[$\nu (^{13}CO)/\nu (^{12}CO)]^2 [\Delta v (^{13}CO)/\Delta v (^{12}CO)]$ 
[$^{12}$CO]/[$^{13}$CO]  $\tau$($^{13}$CO); where $\nu$ is the
frequency of each transition, $\Delta v$ is the FWHM of the observed
line and [$^{12}$CO]/[$^{13}$CO] = 70 is the isotope ratio (Langer \& Penzias
1990). $T_{exc}$ in these regions are about 
5-10 K, while optical depths $\tau$($^{12}$CO) $\simeq$ 9-24.

\begin{table}
\centering
\caption[]{Gaussian fitting to the CO profiles shown in Fig. 10}
\begin{tabular}{rcccr}
\hline
\hline
Spectrum      & T$_{p}$    &   V$_{CO}$  &  $\Delta$V & $I_{CO}$ \\
              & [K]        &  [\kms]     &  [\kms]    & [K \kms]  \\
\hline
{\bf $^{12}$CO(1-0)} &  5.3       &    --64.1   &  3.3       & 18.5 \\
              &  0.5       &    --52.0   &  2.5       &  1.3\\
              &  5.2       &    --38.1   &  1.0       &  5.9 \\
              &  6.7       &    --34.7   &  3.3       & 23.3 \\
\hline
{\bf $^{12}$CO(2-1)} &  3.8       &    --64.0   &  3.6       & 13.7 \\
              &  0.9       &    --56.5   &  2.1       &  1.9 \\
              &  3.9       &    --38.2   &  1.0       &  3.9 \\
              &  4.8       &    --35.0   &  3.6       & 17.7 \\
\hline
{\bf $^{13}$CO(1-0)} &  1.7       &    --64.3   &  2.5       &  4.4 \\
              &  1.4       &    --38.0   &  1.0       &  1.4 \\
              &  1.5       &    --34.7   &  2.9       &  4.6 \\
\hline
{\bf $^{13}$CO(2-1)} &  0.9       &    --63.9   &  2.6       &  2.6 \\
              &  0.8       &    --38.1   &  0.7       &  0.6 \\
              &  0.9       &    --35.1   &  2.2       &  2.2 \\
\hline
\hline
\end{tabular}
\end{table}

The intensity ratios $^{12}CO)/^{13}CO)$ for the different molecular 
components present in the two positions vary from 4.2 to 5.0 for the 
CO(1-0) lines and from 5.3 to 8.1 for the CO(2-1) lines.
These ratios are similar to that found in Galactic molecular clouds 
(Castets et al. 1990).  

The CO(2-1)/CO(1-0) ratios can give us information on the physical conditions
of the gas. This ratio is more sensitive to gas density than kinetic temperature
(Radford et al. 1991). We are aware, however, that these lines are 
optically thick. In order to extract the physical 
conditions of the observed area one can use the LVG approximation (Castets 
et al. 1990). To determine the $^{12}$CO(2-1)/$^{12}$CO(1-0) ratio, we 
convolve the CO(2-1) spectrum  to the CO(1-0) resolution. The ratios 
obtained range between 0.6 to  0.8 and are consistent with low 
ratio gas (LRG), usually subthermally excited, with kinetic temperatures of 
about 10K and densities n$_{H2} < $  1$\times$10$^3$ \cmtres. Ratios of 
about 1 to 1.2 imply kinetic temperatures of about  20K or larger, and densities
n$_{H2} >$ 1$\times$10$^3$ \cmtres, which are not found (Sakamoto et al. 
1994, Sakamoto et al. 1997).    

The parameters of the ionized gas can be estimated from the radio continuum
image. A flux density $S_{4.85GHz}$ = 2.8 Jy was derived from the  image at
5 GHz.
Adopting an electron temperature of 9000 K, the rms electron density   
$n_e$ obtained from the expressions by Mezger \& Henderson (1967) is in 
the range 4-9 \cmtres. To estimate this value we have assumed a spherical 
\hii\ region with constant electron density.
A rough estimate of the filling factor $f$ can be obtained by taking into
account the larger electron density derived from line ratios ($n_e' < $ 100 
\cmtres, Esteban et al. 1990) or the shell parameters around RCW\,78. 
From the former method, we obtained a filling factor   
$f={(\frac{n_e}{n_e'})}^{0.5}$ = 0.3. Adopting an outer radius 
$R_s$ = 15\arcmin\ = 22 pc (at 5.0 kpc) and a inner radius $R_i$ = 10\arcmin\ 
= 15 pc (at 5.0 kpc), and assuming that 50\% of the surface of the \hii\
region is covered by plasma, $f$ = 0.4. The electron density and ionized 
mass for  $f$ = 0.3-0.4 are in the range  6-15 \cmtres\ and 
(3-5)\por 10$^3$ \msun, respectively. The presence of He\,{\sc ii} was 
considered by multiplying the \hii\ mass by 1.27. For the adopted distance 
uncertainty, the error in the ionized mass is about 40\%.  
 
The number of UV photons necessary to ionize the gas $log \ N_{Ly-c}$ = 49.3.  
The comparison with the UV photon flux emitted by a WN7 is  $log \ N_{*}$ = 
49.4 (Crowther 2007), suggesting that the WR star can maintain the
gas in RCW\,78 ionized.

\subsection{Energetics}

The dynamical age of a wind blown bubble can be estimated as 
$t_\mathrm{d}$ = 0.56$\times$10$^6 R/v_\mathrm{exp}$ yr 
(McCray 1983), where $R$ is the radius of the bubble (pc), 
$v_\mathrm{exp}$ is the expansion velocity (\kms), and the coefficient
is the deceleration parameter. The constant is a mean value between 
the energy and the momentum conserving cases. Adopting $R$ = 22 pc and 
assuming a rather low value of 5 \kms\ for the expansion velocity (since 
no signs of expansion were found from optical and radio data), 
$t_\mathrm{d}$ = 2.4$\times$10$^6$ yr, 
suggesting that the O-type star progenitor of the WR star has 
 contributed in shaping the bubble.

The comparison of the stellar wind mechanical energy $E_\mathrm{w} 
(= L_\mathrm{w}t_\mathrm{d} = \dot MV_\mathrm{w}^2t_\mathrm{d}/2$) released 
by WR\,55 and the kinetic energy of the interstellar bubble $E_\mathrm{k}$ 
= $M_\mathrm{b}v_\mathrm{exp}^2/2$ allow to investigate if the stellar
winds are capable of blowing the bubble. 

To estimate  $E_\mathrm{w}$ we adopt a conservative mass loss rate $\dot{M}
\approx$ (1-3)$\times$10$^{-5}$\,M$_\odot$\,yr$^{-1}$ (Nugis \& Lamers 2000,
Cappa et al. 2004), a terminal wind velocity $V_\mathrm{w}$ = 1100~\kms\ 
(see sect. 1), and a dynamical age $t_\mathrm{d}$ = 5$\times$10$^{5}$ yr,
compatible with the duration of the WR phase. The contribution of the
O-type phase of the star was estimated adopting mean values 
$\dot{M}$ = 2$\times$10$^{-6}$\,M$_\odot$\,yr$^{-1}$, $V_\mathrm{w}$
= 1500~\kms\ (Prinja et~al. 1990; Lamers \& Leitherer 1993), and that 
the stellar wind acted during at least 2$\times$10$^{6}$ yr. The wind 
mechanical energy released during the WR phase of the star is in the range  
(0.6-1.8)$\times$10$^{50}$ erg, while for the O-star phase  a value 
0.9$\times$10$^{50}$ erg is derived. The total energy released through
stellar winds is in the range (1.5-2.7)$\times$10$^{50}$ erg.

To estimate the kinetic energy of the interstellar bubble  $E_\mathrm{k}$ 
we take into account the molecular mass alone since it is two orders of 
magnitude larger than the ionized mass. For the adopted expansion velocity,
we obtained $E_\mathrm{k}$ =  0.3$\times$10$^{50}$ erg. 
The comparison between the kinetic energy of the nebula and the 
mechanical energy of the stellar wind is compatible with a interstellar 
bubble interpretation.

\subsection{Star formation process}

The presence of star formation activity in the environs of the nebula
suggests that it may have been triggered by the expansion of the bubble.
Indeed, active star forming regions are present at the periphery of many \hii\ 
regions, where stellar fomation activity is  triggered by the expansion 
of \hii\ regions through the ``collect and collapse'' or the Radiative 
Driven Implosion models (e.g. Deharveng et al. 2005 and references there 
in). 

The "collect and collapse" model indicates that expanding nebulae 
create compressed layers where gas and dust are piled-up between the 
ionization and the shock front. This dense layer fragments forming
molecular cores where new stars born. Whitworth et al. (1994) developed an
analytical model useful to characterize the dense fragments in different
scenarios. 
For the case of stellar wind bubbles, the model predicts the time at 
which the fragmentation occurs $t_{frag}$, the size of the bubble at that 
moment $R_{frag}$, the mass of the fragments $M_{frag}$, and their 
separation along the compressed layer $r_{frag}$. The parameters required 
to derive these quantities are the mechanical luminosity of the wind 
${\it L}_{w}$, the ambient density of the surrounding medium into which 
the bubble expands n$_{0}$, and the isothermal sound speed in the 
shocked gas $a_{s}$, which is supposed to be constant.
 
The analytical expressions derived by Whitworth et al. are:
\begin{equation}
{\it t}_{frag}~[10^{6}~{\rm yr}]~=~0.9~~a_{.2}^{5/8}~~n_{3}^{-1/2}~~L_{37}^{-1/8}
\end{equation}

\begin{equation}
{\it R}_{frag}~[{\rm pc}]~=~9.6~~a_{.2}^{3/8}~~n_{3}^{-1/2}~~L_{37}^{1/8}
\end{equation}


\begin{equation}
{\it M}_{frag}~[{\rm M_{\odot}}]~=~9.8~~a_{.2}^{29/8}~~n_{3}^{-1/2}~~L_{37}^{-1/8}
\end{equation}

\begin{equation}
2 {\it r}_{frag}~[{\rm pc}]~=~0.31~~a_{.2}^{13/8}~~n_{3}^{-1/2}~~L_{37}^{-1/8}
\end{equation}

\noindent where $a_{.2}$ is equal to $a_{s}$ in units of 0.2~\kms, 
$n_{3}$ is the ambient density in units of 1000~\cmtres, and $L_{37}$ 
is $L_{w}$ in units of 10$^{37}$ erg s$^{-1}$. 
$a_{s}$ is an important 
factor to derive the parameters, while  $n_{0}$ has a lower contribution, 
and the dependence with $L_{w}$ is weak. 

Taking into account the stellar wind parameters corresponding to the
main sequence and WR phases of WR\,55, and using the time during which
the wind acted, we estimated $L_{37}$ = (0.20-0.35). Adopting $n_{0}$ 
= 300 \cmtres and a$_s$ = 0.2 \kms, we obtained 
$t_{frag} \ \simeq$ 1.4$\times$10$^6$ yr, 
$R_{frag}$ = 15 pc, $M_{frag}$ = 15 \msun, and $r_{frag}$ = 0.25 pc. 

For this interstellar bubble, the radius $R$ = 22 pc and the dynamical 
age $t_d$ = 2.4$\times 10^{6}$ yr (see Sect. 6.1), and, consequently,
$R_{frag}$ $<$ $R$ and $t_{d} \ >$ $t_{frag}$, suggesting that the bubble is 
old enough for massive molecular fragments to form. Submillimetre continuum data 
may help to detect these fragments. 

Pieces of evidence for the ``collect and collapse'' process are the 
presence of a dense shell surrounding the ionized region and star
formation activity at the periphery of the ionized gas.
The SEST data revealed that molecular gas surrounds the brightest section of 
the nebula (see Sect. 3.2). However, neither signs of strong gas compression 
nor star formation activity are detected at the periphery of this section
of the nebula, as is the case in some \hii\ regions (e.g. Zavagno et al. 
2006, Deharveng et al. 2008).  The NANTEN data showed that most of the 
molecular material appears projected onto the nebula (see Sect. 3.3). 
Although clear evidence of star formation activity seems to be
related to this material, the face-on geometry  
conspires against finding observational evidence in favor of the ``collect 
and collapse'' process.

Stellar formation can also be triggered through the "Radiation-Driven 
Implosion" (RDI) model, theorically developed by Lefloch \& Lazareff (1994). 
Pillars or elephant trumps and cometary globules are observational 
evidences of this process  (e.g. Thompson et al. 2004, Urquhart et al. 
2007). An inspection of the SuperCOSMOS H$\alpha$ does not allow to 
identify these objects. The image at 8 $\mu$m suggests that this kind of 
objects might be present. High resolution molecular line observations 
are needed to investigate the nature of these small features. 

Presently, none of the two processes can be discarded to explain the 
stellar formation activity in the environs of RCW\,78. 

\section{Conclusions}

We have investigated the distribution of the molecular gas related to the
ring nebula RCW\,78 around the WR star HD\,117688 based on observations of 
$^{12}$CO(1-0) and $^{12}$CO(2-1) lines obtained using the SEST. Complementary
molecular data from the NANTEN survey, \hi\ data from the SGPS, and IRAS and 
radio continuum images were also analyzed. 

The analysis of the SEST and NANTEN CO data reveals the existence of molecular
gas having velocities in the range --56 to --33 \kms\ interacting with the
western section of the nebula. The bulk of the emission is concentrated in two 
structures having velocities in the range  --52.5 to --43.5 \kms\ and
--43.5 to --39.5 \kms, coincident with the velocities of the 
H$\alpha$ line.
Based on morphological and kinematical evidences we conclude that the arc-like
structure having velocities in the range --52.5 to --43.5 \kms\ is clearly 
associated with RCW\,78. The feature with velocities from --43.5 to --39.5 
\kms\ is most probably related to the nebula. 
On the contrary, CO emission linked to the  eastern
section of the nebula is difficult to identify.

The radio continuum emission distribution at 4.85 GHz shows a shell-like
feature, which  resembles the H$\alpha$ emission distribution.

The \hi\ emission distribution shows a low emission region coincident in 
position with molecular gas with similar velocities, suggesting
that most of the neutral gas in this region is H$_2$.  In this scenario,
part of the neutral atomic gas encircling the depression might correspond 
to the envelope of the molecular cloud. 

The distribution of the molecular and ionized gas is compatible with a 
scenario in which the ionized gas originated through photodissociation and 
ionization of the parental molecular cloud. 
The shell-like appearance of the ionized gas is suggestive of 
the action of the stellar winds,  which swept-up the surrounding gas
shaping an interstellar bubble. 
The lack of dense gas towards the east favored the expansion of the nebula in 
this direction. The proposed 
scenario also explains the off center location of WR\,55.

As regards stellar formation activity towards RCW\,78, the analysis of the 
IRAS, MSX, and Spitzer point source catalogues revealed the existence
of two active star forming regions liked to the molecular gas 
associated to the nebula. The influence of the expansion of RCW\,78 in 
the onset of star formation in these regions can not be discarded. 

The comparison between the wind mechanical energy released by 
WR\,55 and its massive progenitor and the kinetic energy of the interstellar
bubble indicates that the stellar wind of this star is capable of blowing 
the interstellar bubble. Thus, WR\,55 is not only responsible for the 
ionization of the gas in the nebula but for the creation of the
interstellar bubble. 

\begin{acknowledgements}
C.E.C. acknowledges the kind hospitality of Dr. M. Rubio and her family during 
her stays in Santiago, Chile. We are grateful to Dr. N. Mizuno for
providing us the NANTEN data. We thank the referee for her/his comments and 
suggestions that improved this presentation.
This project was partially financed by CONICET of Argentina under project 
PIP 5886/05,  UNLP under project 11/G093, and ANPCyT under project PICT 
14018/03. M.R. is supported by the Chilean {\sl Center for Astrophysics}
FONDAP No. 15010003. M.R. and C.E.C. wishes to
acknowledge support from FONDECYT (Chile) grant No. 1080335.

This research has made use of the NASA/ IPAC Infrared Science Archive,
which is operated by the Jet Propulsion Laboratory, California Institute
of Technology, under contract with the National Aeronautics and Space
Administration.
 This publication makes use of data products from the Two Micron All Sky
Survey, which is a joint project of the University of Massachusetts and
the Infrared Processing and Analysis Center/California Institute of
Technology, funded by the National Aeronautics and Space Administration
and the National Science Foundation.
  The MSX mission is sponsored by the Ballistic Missile Defense
Organization (BMDO).
We acknowledge the use of NASA's SkyView facility
    (http://skyview.gsfc.nasa.gov) located at NASA Goddard
    Space Flight Center.
This research has made use of the SIMBAD database, operated at CDS,
Strasbourg, France.

\end{acknowledgements}

\end{document}